\documentclass[12pt]{article}

\usepackage{epsf}
\usepackage{amsmath}
\usepackage{graphics}
\usepackage{amsfonts}
\usepackage{amssymb}
\usepackage{latexsym}
\usepackage{color}
\input{colordvi.tex}

\renewcommand{\thefootnote}{\#\arabic{footnote}}

\begin{document}
\setcounter{footnote}{0}

\begin{titlepage}

\begin{center}

\hfill TU-826\\
\hfill October 2008\\

\vskip .5in

{\Large \bf
Non-Gaussianity and Baryonic Isocurvature Fluctuations in the Curvaton
Scenario
}

\vskip .45in

{\large
Takeo Moroi$\,^1$ and
Tomo Takahashi$\,^2$ \\
}

\vskip .45in

{\em
$^1$
Department of Physics, Tohoku University, Sendai 980-8578, Japan}

\vspace{5mm}

{\em
$^2$
Department of Physics, Saga University, Saga 840-8502, Japan
}

\end{center}

\vskip .4in

\begin{abstract}

  We discuss non-Gaussianity and baryonic isocurvature fluctuations in
  the curvaton scenario, assuming that the baryon asymmetry of the
  universe originates only from the decay products of the inflaton.
  When large non-Gaussianity is realized in such a scenario,
  non-vanishing baryonic isocurvature fluctuations can also be
  generated unless the baryogenesis occurs after the decay of the
  curvaton.  We calculate the non-linearity parameter $f_{\rm NL}$ and
  the baryonic isocurvature fluctuations, taking account of the
  primordial fluctuations of both the inflaton and the curvaton.  We
  show that, although current constraints on isocurvature fluctuations
  are severe, the non-linearity parameter can be large as $f_{\rm
    NL}\sim\mathcal{O}(10-100)$ without conflicting with the
  constraints.

\end{abstract}
\end{titlepage}

\renewcommand{\thepage}{\arabic{page}}
\setcounter{page}{1}
\renewcommand{\thefootnote}{\#\arabic{footnote}}

\section{Introduction}

Current cosmological observations are now very precise to give us much
information about the early universe. In particular, from observations
of the cosmic microwave background (CMB) such as WMAP
\cite{Komatsu:2008hk,Dunkley:2008ie}, we can probe the physics of the
early universe through the nature of primordial fluctuations, which
are usually characterized by the amplitude of the fluctuations and its
scale dependence as well as by a possible contribution of gravity
waves.  Recently in addition to these quantities, non-Gaussianity has
been attracting much attention since it provides information about
different aspects of the physics of the early universe.
Non-Gaussianity is usually quantified by the so-called non-linearity
parameter $f_{\rm NL}$ and the recent constraint on this quantity from
WMAP5 is \cite{Komatsu:2008hk}\footnote
{Here $f_{\rm NL}$ represents the so-called ``local type''
  non-Gaussianity.  In this letter, we only consider non-Gaussianity
  of this type.}
\begin{eqnarray}
  -9 < f_{\rm NL} < 111 ~~~ (95\ \% \mbox{ C.L}).
\end{eqnarray}
Although the data is still consistent with Gaussian fluctuation, which
corresponds to $f_{\rm NL} = 0$, the central value is away from zero
as $f_{\rm NL} \sim 50$.  If future observations confirm such non-zero
large value of $f_{\rm NL}$, it will give very important implications
to the scenario of the early universe; a simple model of inflation
would be excluded since fluctuations from the inflaton are almost
Gaussian.  In such a case, some mechanism is needed to generate large
non-Gaussian fluctuations.

Importantly, the simple inflationary scenario is not the only
possibility to generate the density fluctuations.  In particular, from
the viewpoint of particle physics, there may exist a scalar field
other than the inflaton, i.e., so-called the curvaton
\cite{Enqvist:2001zp, Lyth:2001nq, Moroi:2001ct}, which acquires
primordial fluctuation and generate the present density fluctuations
although it is a sub-dominant component during inflation.  With the
curvaton, large non-Gaussianity can also be generated
\cite{Lyth:2002my,curvnongauss},\footnote
{Some other models generating large non-Gaussianity have also been
  discussed, such as the modulated reheating scenario
  \cite{modulated}.}
and the curvaton scenario seems attractive in the light of
constructing a successful model of generating large non-Gaussianity.

In order to generate large non-Gaussianity with the curvaton, it is
necessary that the energy density of the curvaton at the time of its
decay should be much smaller than that of the dominant component of
the universe (which is expected to be from the inflaton).  Density
fluctuations generated in such a scenario are (almost)
scale-invariant, and hence can be consistent with the observations if
the fluctuations of all the components are adiabatic.  However, in
such a scenario, there exist entropy fluctuations between components
from the inflaton and those from the curvaton.  In order to have
vanishing isocurvature fluctuations, it is necessary to generate
baryon asymmetry and cold dark matter (CDM) at low-temperature
universe after the decay of the curvaton.  Although there are several
possibilities that the dark-matter density is determined at a
relatively low temperature, such as axion dark matter, the lightest
superparticle dark matter and so on, baryogenesis at a low temperature
may be challenging.  Indeed, for some of the scenarios of
baryogenesis, like the thermal \cite{Fukugita:1986hr} and non-thermal
\cite{nonthermalleptogen} leptogenesis, a relatively high cosmic
temperature is required.\footnote
{However, the electroweak baryogenesis \cite{Nelson:1991ab} may be a
  possibility.  Another possible model is Affleck-Dine baryogenesis
  \cite{Affleck:1984fy} and the issues of non-Gaussianity in the model
  is discussed in \cite{Kawasaki:2008jy}.}
This fact indicates that, to realize large non-Gaussianity, it may be
necessary to generate the baryon asymmetry of the universe before the
decay of the curvaton.  Since the size of baryonic isocurvature
fluctuations is now severely constrained by observations, such
constraints are important in considering baryogenesis in the curvaton
scenario generating large non-Gaussianity.  In particular, in the
simplest curvaton scenario where the cosmic density fluctuations are
totally from the primordial fluctuation of the curvaton, large
non-Gaussianity cannot be generated without conflicting with the
constraints from the baryonic isocurvature fluctuations if the baryon
asymmetry is solely from the decay product of the inflaton.  This
argument excludes some of the scenarios of baryogenesis which require
high cosmic temperature in the curvaton scenario.

However, in the mixed fluctuation scenario where fluctuations from the
inflaton and the curvaton both contribute to cosmic density
fluctuations \cite{mixed,Ichikawa:2008iq}, the situation changes; even
if the baryon number of the universe originates only from the decay
products of the inflaton, the amplitude of isocurvature fluctuations
relative to adiabatic ones can be suppressed.  Thus, in such a case,
large non-Gaussianity may be generated without conflicting with the
constraint on the isocurvature fluctuations.

In this letter, we discuss non-Gaussianity and baryonic isocurvature
fluctuations in the curvaton scenario where only the decay products of
the inflaton are responsible for the baryogenesis.  We pay particular
attention to the question how large $f_{\rm NL}$ can be while imposing
the isocurvature constraints.  We will see that $f_{\rm
  NL}\sim\mathcal{O}(10-100)$ can be realized even if severe
constraints on isocurvature fluctuations are imposed.

\section{Scenario and Formalism}
\label{sec:scenario}

We first describe the scenario we consider in this letter and
summarize formulae to calculate density fluctuations and its
non-Gaussianity.

Here, we consider the curvaton scenario where fluctuations from the
inflaton also contribute to the total curvature fluctuations along
with that from the curvaton.  In addition, we assume that only the
decay products of the inflaton (not those of curvaton) are responsible
for the baryon asymmetry of the universe; the possibilities include a
baryogenesis in the thermal bath before the decay of the curvaton and
that in association with the decay of the inflaton.  We also assume
that the density of CDM is determined after the decay of the curvaton
so that there is no CDM isocurvature fluctuations.  We can easily
apply our results to the case where CDM originates only from the decay
products of the inflaton.

In our study, we adopt the following form for the scalar-field
potential:
\begin{eqnarray}
\label{eq:V}
V (\phi, \sigma) = V(\phi) + \frac{1}{2} m_\sigma^2 \sigma^2,
\end{eqnarray}
where $\phi$ and $\sigma$ represent the inflaton and curvaton fields,
respectively, $V(\phi)$ is the potential for the inflaton, and
$m_\sigma$ is the mass of the curvaton.  (We assume a quadratic
potential for the curvaton.)  In general, the curvature fluctuations
from the inflaton depend on the inflaton potential.  However, for the
purpose of the following discussion, we only need to specify the value
of the slow-roll parameter $\epsilon$, which is defined as
\begin{eqnarray}
  \label{eq:def_epsilon}
  \epsilon = \frac{1}{2} M_{\rm pl}^2 
  \left( \frac{V_\phi}{V}  \right)^2,
\end{eqnarray}
where $V_\phi\equiv\partial V/\partial\phi$, and $M_{\rm pl}\simeq
2.4\times 10^{18}\ {\rm GeV}$ is the reduced Planck scale.\footnote
{In the following, we assume that the primordial fluctuations are
  (almost) scale-invariant.}
We also assume that the mass of the curvaton is small and the curvaton
acquires primordial fluctuation during the inflation.  (We denote the
initial amplitude of the curvaton as $\sigma_\ast$; hereafter, the
subscript ``$\ast$'' is for quantities at the time of the horizon
exit.)

After the inflation, the inflaton begins to oscillate around the
minimum of its potential and then decays.  We denote the decay rate of
the inflaton as $\Gamma_\phi$ and define the reheating temperature
$T_R$ as
\begin{eqnarray}
  T_R = 
  \left( \frac{10}{g_{\rm SM} \pi^2} M_{\rm pl}^2 \Gamma_\phi^2 
  \right)^{1/4},
\end{eqnarray}
where we use $g_{\rm SM}=106.75$ as the effective number of the
massless degrees of freedom.  As the universe expands, the expansion
rate of the universe $H$ becomes comparable to $m_\sigma$ and the
curvaton starts to oscillate.  Then, when the expansion rate becomes
comparable to the decay rate of $\sigma$, which is denoted as
$\Gamma_\sigma$ (and is related to the lifetime of the curvaton as
$\tau_\sigma=\Gamma_\sigma^{-1}$), the curvaton decays.  The start of
the oscillation and the decay may occur before or after the reheating
due to the inflaton decay, depending on the values of $m_\sigma$ and
$\Gamma_\sigma$.

Assuming that the potential of the inflaton is well approximated by a
quadratic one around its minimum, its energy density behaves as that
of matter for the period of the inflaton oscillation.  Then, denoting
the energy densities of radiation components from the decays of $\phi$
and $\sigma$ as $\rho_{\gamma_\phi}$ and $\rho_{\gamma_\sigma}$,
respectively, evolutions of these variables (as well as those of the
energy density of the inflaton field and the curvaton amplitude) are
governed by
\begin{eqnarray}
  &&\dot{\rho}_{\gamma \phi} + 4 H \rho_{\gamma_\phi}  
  = \Gamma_\phi \rho_\phi, 
  \label{eq:background1} \\
  &&\dot{\rho}_{\gamma \sigma} + 4 H \rho_{\gamma_\sigma}  
  =  \Gamma_\sigma \dot{\sigma}^2,   \label{eq:background2} \\
  &&\dot{\rho}_\phi + 3 H \rho_\phi  =  -\Gamma_\phi \rho_\phi, 
  \label{eq:background3}\\
  &&\ddot{\sigma} +( 3 H  + \Gamma_\sigma) \dot{\sigma} 
  + m_\sigma^2 \sigma =  0, 
  \label{eq:background4}
\end{eqnarray}
where the dot represents derivative with respect to the cosmic time
and  $\rho_i$ indicates the background energy
density of the component $i$.  Notice that, when $H\ll m_\sigma$,
Eqs.\ \eqref{eq:background2} and \eqref{eq:background4} can be well
approximated by
\begin{eqnarray}
  &&\dot{\rho}_{\gamma_\sigma} + 4 H \rho_{\gamma_\sigma}  
  =  \Gamma_\sigma \rho_\sigma,\\
  &&\dot{\rho}_\sigma + 3 H \rho_\sigma  =  -\Gamma_\sigma \rho_\sigma, 
\end{eqnarray}
respectively, where $\rho_\sigma$ is the energy density of the
curvaton field.  In addition, it should be noted that the total
radiation energy density is given by
\begin{eqnarray}
  \rho_{\gamma} = \rho_{\gamma_\phi} + \rho_{\gamma_\sigma}.
\end{eqnarray}

In our analysis, we use the $\delta N$ formalism \cite{deltaN} to
calculate the perturbations.  Then, perturbation variables are
obtained by evaluating the number of $e$-folds from some time during
inflation to the time after the curvaton decay as a function of model
parameters such as $\Gamma_\phi, \Gamma_\sigma, m_\sigma$ and
$\sigma_\ast$.  Here, let us briefly summarize the resultant formulae
of the density fluctuations and non-linearity parameter $f_{\rm NL}$.

With the $\delta N$ formalism, if there is no isocurvature
fluctuation, the curvature fluctuation originating from scalar-field
fluctuations is given by
\begin{eqnarray}
  \zeta^{\rm (adi)} = N_a \delta \varphi^a_\ast
  +\frac{1}{2}N_{ab} \delta \varphi_*^a \delta \varphi_*^b
  + \cdots,
  \label{eq:zeta_expansion}
\end{eqnarray}
where $N$ is the number of $e$-folds, $\delta\varphi^a_*$ is the
primordial fluctuation of the scalar field $\varphi^a$, and
\begin{eqnarray}
  N_a \equiv \frac{\partial N}{\partial \varphi^a}, ~~~
  N_{ab} \equiv 
  \frac{\partial^2 N}{\partial \varphi^a \partial \varphi^b},
\end{eqnarray}
with $\varphi^a=\phi$ and $\sigma$ in our case.

In the following, we consider the scenario where fluctuations of the
inflaton and the curvaton are both responsible for cosmic density
fluctuations.  For simplicity, we assume that these fields are
uncorrelated.  Then, the curvature perturbations originating from
these scalar fields can be calculated separately.  The curvature
fluctuation from the inflaton, which we denote $\zeta^{\rm
  (adi)}_\phi$, is given by
\begin{eqnarray}
  \label{eq:zeta_inf}
  \zeta^{\rm (adi)}_\phi 
  \simeq \frac{1}{\sqrt{2\epsilon} M_{\rm pl} } \delta \phi_\ast 
  + \frac{1}{2} \left( 1 - \frac{\eta}{2\epsilon} \right) 
  \delta \phi_\ast^2,
\end{eqnarray}
where $\eta\equiv M_{\rm pl}^2V_{\phi\phi}/V$, and we have used the
slow-roll approximation.  (We neglect terms of the order of
$\delta\phi_\ast^3$ which are irrelevant for our discussion.)  In
addition, the curvaton contribution to $\zeta^{\rm (adi)}$, which we
denote $\zeta^{\rm (adi)}_\sigma$, is expressed as
\begin{eqnarray}
  \label{eq:zeta_sigma}
  \zeta^{\rm (adi)}_\sigma =  N_\sigma \delta \sigma_\ast + 
  \frac{1}{2} N_{\sigma\sigma} \delta \sigma_\ast^2.
\end{eqnarray} 
We solve the set of equations Eqs.\ \eqref{eq:background1} --
\eqref{eq:background4} with parameters $\Gamma_\phi, \Gamma_\sigma,
m_\sigma$ and $\sigma_\ast$ being fixed, then we determine $N_\sigma$
and $N_{\sigma\sigma}$ to obtain $\zeta_\sigma^{\rm (adi)}$.  For the
important case where
$[\rho_{\gamma_\sigma}/\rho_\gamma]_{t\gg\tau_\sigma}\ll 1$ (and
$m_\sigma\ll\Gamma_\phi$), in which large non-Gaussianity can be
generated in the curvaton scenario, the relation
$4N(\sigma_\ast)=[\rho_{\gamma_\sigma}/\rho_\gamma]_{t\gg\tau_\sigma}$
holds and the $\sigma_\ast$ dependence of $N$ is well approximated as
\cite{Ichikawa:2008iq}
\begin{eqnarray}
  N(\sigma_\ast) \simeq
  \frac{1}{3 \sqrt{2\pi}}\Gamma^2 (5/4)
  \frac{\sigma_\ast^2}{M_{\rm pl}^2\sqrt{\Gamma_\sigma/m_\sigma}}
  ~~~:~~~\mbox{for }
  [\rho_{\gamma_\sigma}/\rho_\gamma]_{t\gg\tau_\sigma}\ll 1,
  \label{eq:Nsmallsigma}
\end{eqnarray}
while, for the case where the curvaton eventually dominates the
universe, $[\rho_{\gamma_\sigma}/\rho_\gamma]_{t\gg\tau_\sigma}\simeq
1$, it is given by
\begin{eqnarray}
  N(\sigma_\ast) \simeq \frac{2}{3} \ln \sigma_\ast
  ~~~:~~~\mbox{for }
  [\rho_{\gamma_\sigma}/\rho_\gamma]_{t\gg\tau_\sigma}\simeq 1.
  \label{eq:Nlargesigma}
\end{eqnarray}

Now, we discuss the entropy fluctuation between baryon and radiation,
which is given up to the second order by
\begin{eqnarray}
  \label{eq:S_b_gamma}
  S_{b \gamma} 
  \equiv    
  \frac{\delta \rho_b}{\rho_b} 
  - \frac{1}{2} \left( \frac{\delta \rho_b}{\rho_b}  \right)^2
  - 
  \frac{3}{4} \left[ 
  \frac{\delta \rho_\gamma}{\rho_\gamma} 
  - \frac{1}{2} \left( \frac{\delta \rho_\gamma}{\rho_\gamma} \right)^2
  \right].
\end{eqnarray}
Here, $\delta\rho_i$ denotes the energy-density fluctuation of the
component $i$ on the uniform density slicing.  We evaluate $S_{b
  \gamma}$ when $\tau_\sigma\ll t\ll t_{\rm eq}$ with $t_{\rm eq}$
being the time of the radiation-matter equality.  In the present
scenario, there exist two sources of radiation: one originating from
the inflaton and the other from the curvaton.  We treat them
separately and write $\delta \rho_\gamma / \rho_\gamma$ as
\begin{eqnarray}
\frac{\delta \rho_\gamma}{\rho_\gamma} 
=
  \frac{\delta \rho_{\gamma_\phi} + \delta \rho_{\gamma_\sigma}}
  {\rho_{\gamma_\phi} + \rho_{\gamma_\sigma}}.
  \label{eq:delta_gamma}
\end{eqnarray}
Since the baryon asymmetry and $\gamma_\phi$ has the same source
(i.e., the inflaton), the adiabatic relation holds between these two
components.  Furthermore, as will be discussed later, large
non-Gaussianity can be generated when $ \left[ \rho_{\gamma_\sigma} /
  \rho_{\gamma} \right]_{t\gg\tau_\sigma} \ll 1$, thus we concentrate
on such a case.  Neglecting the terms which are second or higher order
in $ \left[ \rho_{\gamma_\sigma} / \rho_{\gamma}
\right]_{t\gg\tau_\sigma}$, and using the adiabatic relation between
the baryon and $\gamma_\phi$, we obtain
\begin{eqnarray}
  S_{b \gamma} 
  \simeq
    - \frac{3}{4} 
   \left[
    \frac{\rho_{\gamma_\sigma}}{\rho_{\gamma}}
    \right]_{t\gg\tau_\sigma}
  \frac{\delta \rho_{\gamma_\sigma}}{\rho_{\gamma_\sigma}},
\end{eqnarray}
where terms of the order of $(\delta \rho_{\gamma_\sigma} /
\rho_{\gamma_\sigma})^2$ vanish.  (Here and hereafter, it should be
understood that $[\rho_{\gamma_\sigma}/\rho_\gamma]_{t\gg\tau_\sigma}$
is equal to $4N(\sigma_\ast)$, and is proportional to
$\sigma_\ast^2$.)  Notice that, compared to
$\delta\rho_{\gamma_\sigma}/\rho_{\gamma_\sigma}$,
$\delta\rho_{\gamma_\phi}/\rho_{\gamma_\phi}$ is of the order of
$\left[ \rho_{\gamma_\sigma} / \rho_{\gamma}
\right]_{t\gg\tau_\sigma}$ and hence its contribution is irrelevant in
the present discussion.

When
$\left[\rho_{\gamma_\sigma}/\rho_{\gamma}\right]_{t\gg\tau_\sigma} \ll
1$, the cosmic expansion is solely determined by radiation from the
inflaton, and we can neglect the effect of $\gamma_\sigma$ on the
background evolution.  Then, $\rho_{\gamma_\sigma}$ is proportional to
$\sigma_\ast^2$ and
\begin{eqnarray}
  \frac{\delta \rho_{\gamma_\sigma}}{\rho_{\gamma_\sigma}} 
  =
  2
  \left( 
    \frac{\delta\sigma_\ast}{\sigma_\ast}
    + \frac{1}{2} \frac{\delta\sigma_\ast^2}{\sigma_\ast^2}
  \right),
  \label{eq:delta_rho_sigma}
\end{eqnarray}
which results in
\begin{eqnarray}
  S_{b \gamma}
  =
  -\frac{3}{2}
  \left[
    \frac{\rho_{\gamma_\sigma}}{\rho_{\gamma}}
  \right]_{t\gg\tau_\sigma}
  \left( 
    \frac{\delta\sigma_\ast}{\sigma_\ast}
    + \frac{1}{2} \frac{\delta\sigma_\ast^2}{\sigma_\ast^2}
  \right).
  \label{eq:S_bgamma2}
\end{eqnarray}

In discussing the non-Gaussianity in the present framework, it should
be noted that the isocurvature fluctuations also generate curvature
perturbation.  Indeed, using the relation $3\zeta^{\rm
  (iso)}=S_{m\gamma}$ (with $S_{m\gamma}$ being the entropy
fluctuation between the total matter and radiation), which holds in
the matter-dominated universe, we obtain the isocurvature contribution
to the curvature fluctuation as
\begin{eqnarray}
  \zeta^{\rm (iso)}_\sigma = -\frac{1}{2} 
  \frac{\Omega_b}{\Omega_m} 
  \left[
    \frac{\rho_{\gamma_\sigma}}{\rho_{\gamma}}
  \right]_{t\gg\tau_\sigma}
  \left( 
    \frac{\delta\sigma_\ast}{\sigma_\ast}
    + \frac{1}{2} \frac{\delta\sigma_\ast^2}{\sigma_\ast^2}
  \right),
  \label{eq:zeta_iso}
\end{eqnarray}
where $\Omega_b$ and $\Omega_m$ are density parameters of the total
matter and baryon, respectively.  In our numerical analysis, we use
$\Omega_b/\Omega_m\simeq 0.17$ \cite{Komatsu:2008hk}.

Since the fluctuations of the inflaton and the curvaton are assumed to
be uncorrelated, the amplitude of the total curvature fluctuation
$\zeta$ in matter dominated epoch is given by
\begin{eqnarray}
  \zeta = \zeta_\phi + \zeta_\sigma,
\end{eqnarray}
where $\zeta_\phi=\zeta^{\rm (adi)}_\phi$ and $\zeta_\sigma=\zeta^{\rm
  (adi)}_\sigma+\zeta^{\rm (iso)}_\sigma$.  From the observations of
the cosmic density fluctuations, the size of $\zeta$ is constrained.
In our study, we determine the amplitudes of the primordial
scalar-field fluctuations 
so that $\zeta$ becomes consistent with the
observed value.

Non-Gaussianity of fluctuations is usually quantified with 
higher order statistics such as bispectrum.
 Here, we consider the bispectrum of
$\zeta$:
\begin{eqnarray}
  \langle \zeta_{\vec k_1} \zeta_{\vec k_2} \zeta_{\vec k_3} \rangle
  &=&
  {(2\pi)}^3
  \delta ({\vec k_1}+{\vec k_2}+{\vec k_3}) B_\zeta (k_1,k_2,k_3).
  \label{eq:bi}
\end{eqnarray}
Then, $B_\zeta (k_1,k_2,k_3)$ is obtained as
\begin{eqnarray}
  B_\zeta (k_1,k_2,k_3) = 
  \left( 
    \tilde{N}_\sigma^2 
    + \tilde{N}_{\sigma\sigma}^2 \Delta_{\delta\sigma}^2 \ln k_{\rm min} L
  \right) 
  \tilde{N}_{\sigma\sigma}
  \left[
    P_{\delta\sigma} (k_1) P_{\delta\sigma} (k_2)
    + (\mbox{2 perms.})
  \right],
  \label{B123}
\end{eqnarray}
where
\begin{eqnarray}
  \tilde{N}_\sigma &=& 
  N_\sigma - \frac{1}{2} 
  \frac{\Omega_b}{\Omega_m} 
  \left[
    \frac{\rho_{\gamma_\sigma}}{\rho_{\gamma}}
  \right]_{t\gg\tau_\sigma} 
  \sigma_\ast^{-1},
 \label{eq:tildeN_s} \\
  \tilde{N}_{\sigma\sigma} &=& 
  N_{\sigma\sigma} - \frac{1}{2} 
  \frac{\Omega_b}{\Omega_m} 
  \left[
    \frac{\rho_{\gamma_\sigma}}{\rho_{\gamma}}
  \right]_{t\gg\tau_\sigma} 
  \sigma_\ast^{-2}.
   \label{eq:tildeN_ss} 
\end{eqnarray}
Here, $P_X(k)$ denotes the power spectrum of the variable $X$ defined
as
\begin{eqnarray}
  \langle X_{\vec k_1} X_{\vec k_2} \rangle
  =
  {(2\pi)}^3 \delta ({\vec k_1}+{\vec k_2}) P_X (k_1),
\end{eqnarray}
and is related to $\Delta_X^2$ as
\begin{eqnarray}
  P_X (k) = \frac{2 \pi^2}{k^3} \Delta_X^2.
\end{eqnarray}
Assuming that the curvaton fluctuation is due to the quantum
fluctuation during inflation, we obtain
\begin{eqnarray}
  \Delta_{\delta\sigma}^2 = \left( \frac{H_*}{2\pi} \right)^2,
\end{eqnarray}
where $H_\ast$ is the expansion rate during inflation and 
we assume $ \Delta_{\delta\sigma}^2 $ to be scale-invariant.

 In deriving Eq.\
\eqref{B123}, following \cite{Lyth:1991ub, Boubekeur:2005fj}, we have
regularized the infrared divergence by introducing the the infrared
cutoff parameter $L^{-1}$ and $k_{\rm min}=\mbox{min}(k_1,k_2,k_3)$.  (We
have neglected an $O(1)$ coefficient in front of $\ln k_{\rm min} L$.)
In order to discuss the implication to the cosmological observations,
both $k_{\rm min}$ and $L^{-1}$ are taken to be the cosmological
scale.  Since the scale dependence from $\ln k_{\rm min} L$ is rather
weak, we approximate $\ln k_{\rm min} L=1$ in our numerical analysis.

As mentioned in the introduction, the non-linearity parameter $f_{\rm
  NL}$ is often used to characterize non-Gaussianity of fluctuations,
which is given by\footnote{
Since there exists the contribution from isocurvature fluctuations in
this scenario, $f_{\rm NL}$ here is not the same as the one for the
case only with adiabatic perturbations.  However, as shown in
Eq.~\eqref{eq:zeta_iso}, the isocurvature contribution is suppressed
by the factor $\Omega_b / \Omega_m$.  In fact, because of the
difference between the transfer function for the adiabatic
contribution and that for the isocurvature one, the bispectrum is
enhanced at the Sachs-Wolfe plateau.  In the present case, the
enhancement factor for the corresponding term 
is roughly estimated to be $\sim 2^{2/3}$ for large-scale
fluctuations \cite{Kawasaki:2008sn}.  Even if the second terms in
Eqs.~\eqref{eq:tildeN_s} and \eqref{eq:tildeN_ss} are multiplied by
this factor, the contribution from the isocurvature fluctuations is
still sub-dominant.
 For the details of non-Gaussianity from isocurvature fluctuations, see
  \cite{Kawasaki:2008sn}.
  }
\begin{eqnarray}
  B_\zeta (k_1,k_2,k_3)
  =
  \frac{6}{5} f_{\rm NL}
  \left[
    P_\zeta (k_1) P_\zeta (k_2) + (\mbox{2 perms.})
  \right].
  \label{eq:def_f_NL}
\end{eqnarray}
Here $P_\zeta (k)$ is the the power spectrum of $\zeta$ defined as
\begin{eqnarray}
  \label{eq:power}
  \langle \zeta_{\vec k_1} \zeta_{\vec k_2} \rangle
  =
  {(2\pi)}^3 \delta ({\vec k_1}+{\vec k_2}) P_\zeta (k_1).
\end{eqnarray}
In the present scenario, $P_\zeta (k)$ becomes
\begin{eqnarray}
  P_\zeta (k) = 
  \frac{2 \pi^2}{k^3} \Delta_\zeta^2
  = \frac{2 \pi^2}{k^3}
  \left[
    \frac{1}{2\epsilon M_{\rm pl}^2} \Delta_{\delta\phi}^2
    + \left( 
      \tilde{N}_\sigma^2 
      + \tilde{N}_{\sigma\sigma}^2 \Delta_{\delta\sigma}^2 \ln k L
    \right) \Delta_{\delta\sigma}^2
  \right].
\end{eqnarray}
Then, the non-linearity parameter $f_{\rm NL}$ is given by
\begin{eqnarray}
  \frac{6}{5} f_{\rm NL} 
  =
  \frac{(2 \epsilon -\eta) 
    + 4\epsilon^2 M_{\rm pl}^4 
    (\tilde{N}_\sigma^2 
    + \tilde{N}_{\sigma\sigma}^2 \Delta_{\delta\sigma}^2 \ln k_{\rm min} L)
    \tilde{N}_{\sigma\sigma}}
  {\left[ 1+ 2 \epsilon M_{\rm pl}^2 (\tilde{N}_\sigma^2 
      + \tilde{N}_{\sigma\sigma}^2 \Delta_{\delta\sigma}^2 \ln k_{\rm min} L)
    \right]^2},
  \label{eq:f_NL}
\end{eqnarray}
where we have used the relation
$\Delta_{\delta\phi}^2=\Delta_{\delta\sigma}^2$.  Since we are
interested in the case where the non-linearity parameter becomes
large, we neglect $(2 \epsilon -\eta)$ in the numerator of
\eqref{eq:f_NL} in the following analysis.  We can find approximated
formulae of $\tilde{N}_\sigma$ and $\tilde{N}_{\sigma\sigma}$ for the
most important case of
$[\rho_{\gamma_\sigma}/\rho_\gamma]_{t\gg\tau_\sigma}\ll 1$, for which
$f_{\rm NL}\gg 1$ may be realized.  In such a case, as we have
mentioned, the relation
$4N(\sigma_\ast)=[\rho_{\gamma_\sigma}/\rho_\gamma]_{t\gg\tau_\sigma}$
holds and hence
\begin{eqnarray}
  \tilde{N}_\sigma \simeq
  \left( 1 -  \frac{\Omega_b}{\Omega_m} \right) N_\sigma,
  ~~~
  \tilde{N}_{\sigma\sigma} \simeq 
  \left( 1 -  \frac{\Omega_b}{\Omega_m} \right) 
  N_{\sigma\sigma}
  ~~~:~~~\mbox{for }
  [\rho_{\gamma_\sigma}/\rho_\gamma]_{t\gg\tau_\sigma}\ll 1.
  \label{eq:N_s_N_ss}
\end{eqnarray}

For the sake of the following arguments, we also calculate the power
spectrum of the isocurvature perturbations.  With Eq.\
\eqref{eq:S_bgamma2}, we obtain
\begin{eqnarray}
  \Delta_{S_{b\gamma}}^2 = \frac{9}{4} 
  \left[
    \frac{\rho_{\gamma_\sigma}}{\rho_{\gamma}}
    \right]_{t\gg\tau_\sigma}^2
  \left( \frac{1}{\sigma_*^2}
  + \frac{1}{\sigma_*^4} \Delta_{\delta\sigma}^2 \ln k_{\rm min} L
  \right) \Delta_{\delta\sigma}^2.
  \label{Del_Sbg}
\end{eqnarray}

Before showing the numerical results, we briefly consider the pure
curvaton case where all the cosmic density fluctuations are only from
the curvaton.  In such a case, we obtain
\begin{eqnarray}
  \frac{\Delta_{S_{b\gamma}}}{\Delta_\zeta} \sim
  \frac{N_{\sigma\sigma}}{N_\sigma^2}
  \left[
    \frac{\rho_{\gamma_\sigma}}{\rho_{\gamma}}
    \right]_{t\gg\tau_\sigma},
\end{eqnarray}
and $\Delta_{S_{b\gamma}}$ and $\Delta_\zeta$ are of the same order
irrespective of $\sigma_\ast$.  (See, for example, Eqs.\
\eqref{eq:Nsmallsigma} and \eqref{eq:Nlargesigma}.)  In this case, the
entropy fluctuation is too large to be consistent with the
observations \cite{Moroi:2001ct, Lyth:2002my, Moroi:2002rd,
  Lyth:2003ip}.  Thus when the baryon number is produced at high
temperature, it is difficult to have large non-Gaussianity in the
simplest curvaton paradigm without conflicting with the isocurvature
constraint.  As we will see in the following, the situation changes in
the mixed fluctuation scenario.  In particular, when the curvature
perturbation mainly comes from the inflaton fluctuation and
$[\rho_{\gamma_\sigma}/\rho_\gamma]_{t\gg\tau_\sigma}\ll 1$, large
non-Gaussianity becomes possible without conflicting with the
isocurvature constraint.

\section{Numerical Results}
\label{sec:nonG}

Now, we show our numerical results.  In our analysis, we numerically
solve Eqs.\ \eqref{eq:background1} -- \eqref{eq:background4} and
calculate the number of $e$-folds as a function of $\sigma_\ast$.
Then, we calculate $f_{\rm NL}$ and $S_{b\gamma}$ for various values
of the model parameters.  In the following, we take $m_\sigma = 100\
{\rm GeV}$.

\begin{figure}[t]
  \centerline{\epsfysize=0.35\textheight\epsfbox{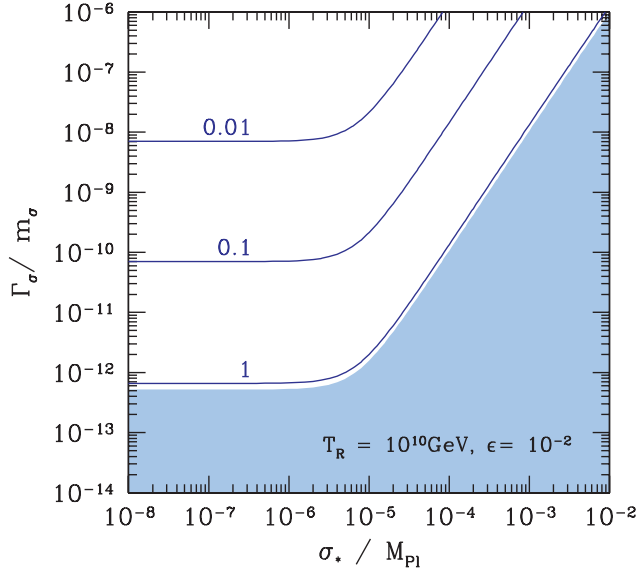}}
  \caption{\small Contours of constant
    $\Delta_{S_{b\gamma}}/\Delta_\zeta$ on $\sigma_\ast$ vs.\
    $\Gamma_\sigma / m_\sigma$ plane for $T_R=10^{10}\ {\rm GeV}$.
    For the slow-roll parameter, we take $\epsilon=10^{-2}$.  Regions
    excluded by the constraints \eqref{eq:SbgBound} are shaded.}
\label{fig:s2zeta}
\end{figure}

First, we show how large the baryonic isocurvature fluctuations
can be.  In Fig.\ \ref{fig:s2zeta}, we show contours of constant
$\Delta_{S_{b\gamma}}/\Delta_\zeta$ on the $\sigma_\ast$ vs.\
$\Gamma_\sigma / m_\sigma$ plane.  Here, we take $T_R=10^{10}\ {\rm
  GeV}$, and $\epsilon=10^{-2}$.\footnote
{In our following numerical analysis, we use $\epsilon=10^{-2}$ and
  $10^{-10}$ for illustrational purposes.  In fact, the value of
  $\epsilon=10^{-2}$ corresponds to the case of quadratic chaotic
  inflation with $N_e=50$, while $\epsilon\sim \mathcal{O}(10^{-10})$
  is realized in some class of new inflation model.}
When $\tilde{N}_\sigma^2\gtrsim\tilde{N}_{\sigma\sigma}^2
\Delta_{\delta\sigma}^2 \ln k_{\rm min} L$, one can see that the
baryonic isocurvature fluctuations are suppressed as $\sigma_\ast$
becomes smaller.  This fact can be easily understood as follows.  When
$\sigma_\ast$ is small enough, the curvature perturbation is dominated
by the inflaton contribution $\zeta_\phi$ and hence is independent of
$\sigma_\ast$.  In addition, in such a case, the ratio
$[\rho_{\gamma_\sigma}/\rho_\gamma]_{t\gg\tau_\sigma}$ is proportional
to $\sigma_\ast^2$ and hence $\Delta_{S_{b\gamma}}/\Delta_\zeta$
becomes smaller as $\sigma_\ast$ decreases as far as the first term in
the parenthesis of Eq.\ \eqref{Del_Sbg} dominates.  On the contrary,
for $\tilde{N}_\sigma^2\lesssim\tilde{N}_{\sigma\sigma}^2
\Delta_{\delta\sigma}^2 \ln k_{\rm min} L$, the isocurvature perturbation 
$\Delta_{S_{b\gamma}}$ is
determined by the second term in Eq.\
\eqref{Del_Sbg}.  Then, $\Delta_{S_{b\gamma}}/\Delta_\zeta$ becomes
insensitive to $\sigma_\ast$, as shown in the figure.

From current cosmological observations, 
baryonic isocurvature fluctuations are severely
constrained; there is no sign of the isocurvature fluctuations in the
observed angular power spectrum of the CMB, and $S_{b\gamma}$ is
consistent with zero.  In our analysis, we adopt the bounds on the
baryonic isocurvature fluctuations obtained from the latest WMAP5
result. We classify the baryonic isocurvature fluctuations into
correlated and uncorrelated parts as
\begin{eqnarray}
  \left[ S_{b \gamma} \right]_{\rm corr} 
  = 
  - \Delta_{S_{b\gamma}}
  \sin \delta,~~~
  \left[ S_{b \gamma} \right]_{\rm uncorr} 
  = 
  \Delta_{S_{b\gamma}}
  \sqrt{1 - \sin^2 \delta},
\end{eqnarray}
where $\sin\delta=\Delta_{\zeta_\sigma}/\Delta_{\zeta}$.  Then, we
adopt the following bounds on the ratios of these isocurvature modes
to $\Delta_\zeta$, reading off the 95\ \% C.L. constraints from the
WMAP5 results \cite{Komatsu:2008hk}
\begin{eqnarray}
  \left[ \frac{S_{b \gamma}}{\Delta_\zeta}  \right]_{\rm corr} 
  > -0.31, ~~~
  \left[ \frac{S_{b \gamma}}{\Delta_\zeta} \right]_{\rm uncorr} 
  < 1.35.
  \label{eq:SbgBound}
\end{eqnarray}
Notice that, in the present scenario, $[S_{b
  \gamma}/\Delta_\zeta]_{\rm corr}$ is negative.\footnote
{The bounds given in \eqref{eq:SbgBound} are separately obtained for
  the cases of totally correlated and totally uncorrelated
  isocurvature fluctuations.  Since the bounds for the case where the
  correlated and uncorrelated isocurvature fluctuations coexist are
  not available, we adopt the constraint \eqref{eq:SbgBound} as
  reference values.  In addition, since the constraint on the
  correlated isocurvature fluctuations are not shown for $[S_{b
    \gamma}/\Delta_\zeta]_{\rm corr}<0$ in \cite{Komatsu:2008hk}, we
  adopt the constraint for the case of $[S_{b
    \gamma}/\Delta_\zeta]_{\rm corr}>0$ to derive the constraint
  \eqref{eq:SbgBound}, assuming that the bound on $|[S_{b
    \gamma}/\Delta_\zeta]_{\rm corr}|$ does not significantly depend
  on the sign.  For the validity of this assumption, see, for example,
  \cite{S_bg}.}
In Fig.\ \ref{fig:s2zeta}, we shaded the region which is excluded by
the above constraints; we found that the constraint on the correlated
one is more stringent.

\begin{figure}[t]
  \centerline{\epsfxsize=\textwidth\epsfbox{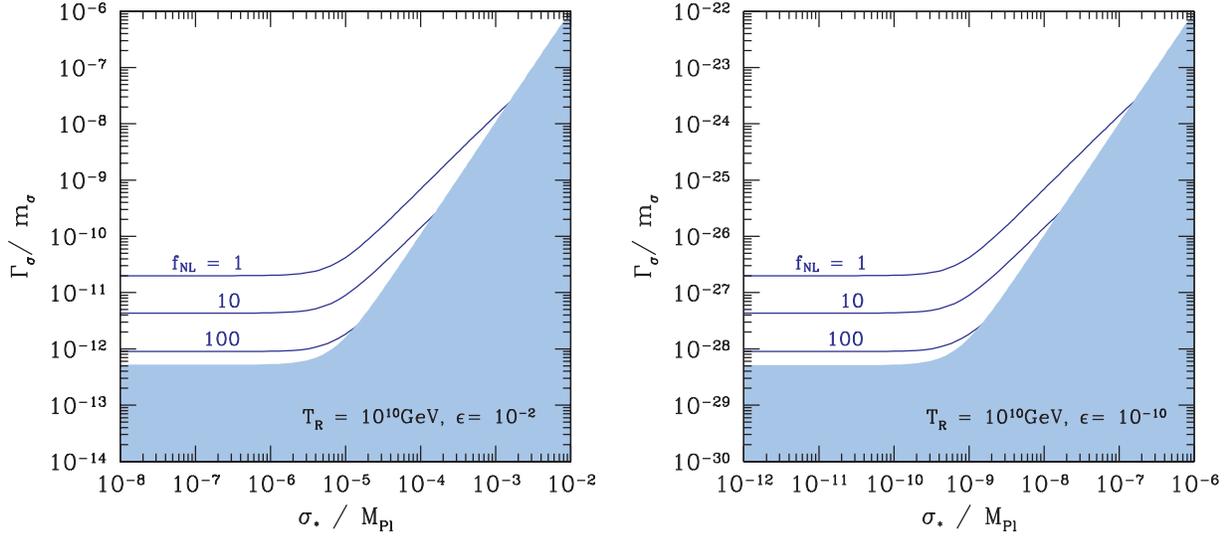}}
  \caption{\small Contours of constant $f_{\rm NL}$ on the
    $\sigma_\ast$ vs.\ $\Gamma_\sigma / m_\sigma$ plane for
    $T_R=10^{10}\ {\rm GeV}$.  ($f_{\rm NL}=1$, $10$, and $100$ from
    the top.)  For the slow-roll parameter, we take $\epsilon=10^{-2}$
    (left) and $10^{-10}$ (right).  Regions excluded by the
    constraints \eqref{eq:SbgBound} are shaded.}
\label{fig:fnlT10}
\end{figure}

It should be noted that, in the allowed region, the curvature
perturbation $\zeta$ is mainly from the inflaton fluctuation.  Then,
$\Delta_{\delta\sigma}^2$ and $\Delta_{\delta\phi}^2$ are related to
$\Delta_{\zeta}^2$, which is known from the observation of the cosmic
density fluctuations, as
\begin{eqnarray}
  \Delta_{\delta\sigma}^2 = \Delta_{\delta\phi}^2 
  = 2\epsilon M_{\rm pl}^2 \Delta_{\zeta}^2.
\end{eqnarray}
Neglecting the scale-dependence, we adopt
$\Delta_{\zeta}^2=2.457\times 10^{-9}$ \cite{Komatsu:2008hk} to
evaluate $f_{\rm NL}$ given in Eq.\ \eqref{eq:f_NL}.

Now we discuss how large $f_{\rm NL}$ can be in the parameter region
consistent with the constraints \eqref{eq:SbgBound}.  In Fig.\
\ref{fig:fnlT10}, taking $T_R=10^{10}$ GeV, we show the contours of
constant $f_{\rm NL}$ for $\epsilon=10^{-2}$ and $10^{-10}$.  With
such a choice of the reheating temperature, the curvaton field starts
to oscillate after the reheating by the inflaton.  If the curvaton
begins to oscillate after the inflaton decay, the number of $e$-folds
after the inflation depends only on the combination of $\Gamma_\sigma
/ m_\sigma$ once $\sigma_*$ is fixed.  Thus, we show our results in
the $\sigma_\ast$ vs.\ $\Gamma_\sigma / m_\sigma$ plane.  (Notice
that, for different values of $T_R$ and $m_\sigma$, the figure is
almost unchanged as far as $m_\sigma\ll\Gamma_\phi$.)  On the same
figure, we shaded the region excluded by the constraints
\eqref{eq:SbgBound}.  We can see that, even after imposing the
isocurvature constraint, a very large value of $f_{\rm NL}$ (i.e.,
$f_{\rm NL}\sim \mathcal{O}(10-100)$) is possible with small enough
$\sigma_\ast$; in such a case, even though the components related to
the curvaton are always sub-dominant, $f_{\rm NL}$ becomes large.  (A
possibility where a sub-dominant component generates large
non-Gaussianity was first considered in \cite{Boubekeur:2005fj}.)  In
addition, when $\sigma_*$ becomes small enough, $f_{\rm NL}$ becomes
insensitive to $\sigma_*$.  This is because, in such a parameter
region, the second term in the first parenthesis of Eq.\ \eqref{B123},
which is independent of $\sigma_*$, dominates.

\begin{figure}[t]
  \centerline{\epsfxsize=\textwidth\epsfbox{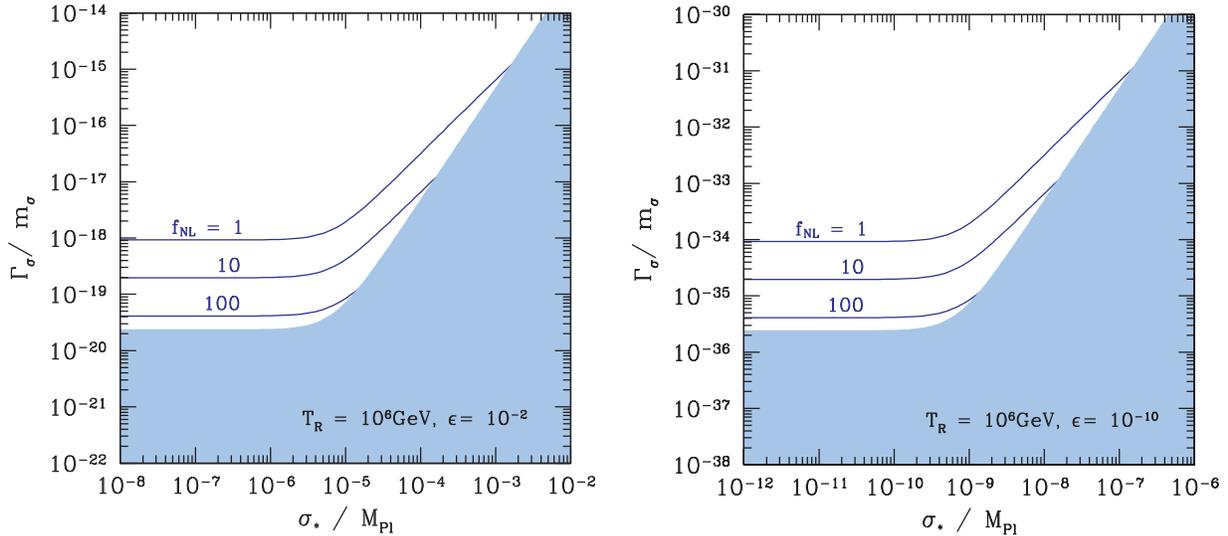}}
  \caption{\small Same as Fig.\ \ref{fig:fnlT10}, except for
    $T_R=10^{6}\ {\rm GeV}$.}
\label{fig:fnlT06}
\end{figure}

Next, let us consider the case where the curvaton decays before the
reheating (i.e., during the inflaton oscillation).  Even in such a
case, large value of $f_{\rm NL}$ can be obtained while satisfying the
isocurvature constraints.  To see this, in Fig.\ \ref{fig:fnlT06}, we
show the contours of constant $f_{\rm NL}$ on $\sigma_\ast$ vs.\
$\Gamma_\sigma / m_\sigma$ plane for the case of $T_R=10^{6}\ {\rm
  GeV}$.  As one can see, in this case, $f_{\rm NL}\sim
\mathcal{O}(10-100)$ can be realized even if we impose the baryonic
isocurvature constraints.

\section{Implications for Scenarios of Baryogenesis}
\label{sec:implication}

So far, we have seen that a large value of the non-linearity parameter
of $f_{\rm NL}\sim \mathcal{O}(10-100)$ can be realized without
conflicting with the baryonic isocurvature constraints, assuming
$T_R$, $\sigma_\ast$, and $\Gamma_\sigma$ as free parameters.  If we
fix the scenario of baryogenesis, however, it is often the case that a
lower bound on $T_R$ is imposed to generate large enough baryon
asymmetry of the universe.  So, finally, we discuss whether large
$f_{\rm NL}$ is possible for several scenarios of baryogenesis.

In order to realize the thermal leptogenesis \cite{Fukugita:1986hr},
$T_R$ should be higher than $10^{9-10}\ {\rm GeV}$
\cite{TR_leptogen}.\footnote
{In the region of interest, the ratio
  $[\rho_{\gamma_\sigma}/\rho_\gamma]_{t\gg\tau_\sigma}$ is much
  smaller than $1$, and hence the entropy production due to the
  curvaton decay is negligible.}
Then, for $m_\sigma=100\ {\rm GeV}$ and $\epsilon=10^{-2}$,
$\sigma_\ast\lesssim 10^{15}\ {\rm GeV}$ and $\Gamma_\sigma\lesssim
10^{-7}\ {\rm GeV}$ are required to realize $f_{\rm NL}=10$ without
conflicting with the constraints \eqref{eq:SbgBound}.  Notice that,
even though $\Gamma_\sigma$ has to be much smaller than $m_\sigma$,
the curvaton decay occurs when the cosmic temperature is $T\gtrsim
10^5\ {\rm GeV}$ for $\Gamma_\sigma=10^{-8}\ {\rm GeV}$, which is well
before the start of the big-bang nucleosynthesis (BBN).\footnote
{Even if the curvaton decays after the BBN, effects of the decay
  products on the abundances of the light elements may not be
  significant because we are particularly interested in the case where
  $[\rho_{\gamma_\sigma}/\rho_\gamma]_{t\gg\tau_\sigma}\ll 1$.  The
  BBN constraints depend on the detail of the decay process (i.e.,
  lifetime, hadronic branching ratio, and so on)
  \cite{Kawasaki:2004qu}, and study of the BBN constraints is beyond
  the scope of this letter.}
If $m_\sigma$ is
larger, $f_{\rm NL}\sim \mathcal{O}(10)$ can be realized with a larger
value of $\Gamma_\sigma$.

Even though the scenario of the thermal leptogenesis is attractive, it
is incompatible with some class of supersymmetric model because, if
$T_R\gtrsim 10^{9-10}\ {\rm GeV}$, overproduction of the gravitino may
occur \cite{Kawasaki:2008qe}.  If a scenario of baryogenesis which
works at a lower temperature is needed, one of the possibilities is
the non-thermal leptogenesis in which right-handed neutrinos are
directly produced by the decay of a scalar condensation (like the
inflaton) \cite{nonthermalleptogen}.  In the non-thermal leptogenesis
scenario, the lower bound on $T_R$ is reduced, and is given by
$T_R\gtrsim 10^6\ {\rm GeV}$ \cite{Ibe:2005jf}.  As one can see, even
with $T_R\sim 10^6\ {\rm GeV}$, $f_{\rm NL}$ can be as large as
$\mathcal{O}(10)$ (or larger) satisfying the baryonic isocurvature
constraints.  In this case, with $m_\sigma=100\ {\rm GeV}$ and
$\epsilon=10^{-2}$, $f_{\rm NL}=10$ requires $\sigma_\ast\lesssim
10^{14}\ {\rm GeV}$ and $\Gamma_\sigma\lesssim 10^{-15}\ {\rm GeV}$,
which corresponds to the decay temperature of the curvaton of $\sim
\mathcal{O}(10\ {\rm GeV})$.

In summary, even if the baryon asymmetry originates only from the
decay products of the inflaton, a large non-Gaussianity of $f_{\rm
  NL}\sim\mathcal{O}(10-100)$ is possible in large class of scenarios
of baryogenesis without conflicting with the observational
constraints.  In such a scenario, however, baryonic isocurvature
fluctuations are inevitably generated and may be just below the
observational bound.  Thus, if a large value of $f_{\rm NL}$ is
confirmed in future observations, it is strongly encouraged to look
for signals of the isocurvature fluctuations to test baryogenesis
models in the curvaton scenario.

{\sl Note added:} While we were finalizing the manuscript,
Ref.~\cite{Langlois:2008vk} appeared on the arXiv, which has some
overlap with our analysis.

{\sl Acknowledgments:} 
One of the authors (T.M.)  acknowledges support from the Aspen Center
for Physics where a part of this work has been done.  This work is
supported in part by the Grant-in-Aid for Scientific Research from the
Ministry of Education, Science, Sports, and Cultures of Japan
No.\,19540255 (T.M.) and No.\,19740145 (T.T.), and the Sumitomo
Foundation (T.T.).

\end{document}